\documentclass[twocolumn,showpacs,prb,footinbib,a4paper,superscriptaddress,floatfix]{revtex4}

\usepackage{amssymb,amsmath}
\usepackage{graphicx}

\newcommand{\GF}{{\rm Green's function}\ }
\newcommand{\GFs}{{\rm Green's functions}\ }
\newcommand{\SE}{{\rm self-energy}\ }
\newcommand{\SEs}{{\rm self-energies}\ }
\newcommand{\NE}{{\rm non-equilibrium}\ }

\begin{document}

\title{Generalization and applicability of the Landauer formula \\ for non-equilibrium
current in the presence of interactions}

\author{H. Ness}
\email{hn506@york.ac.uk}
\author{L. K. Dash}
\author{R. W. Godby}

\affiliation{Department of Physics, University of York, Heslington, York YO10 5DD,
UK}
\affiliation{European Theoretical Spectroscopy Facility (ETSF)}

\date{\today}

\begin{abstract}
  Using non-equilibrium Green's functions (NEGF), we calculate the
  current through an interacting region connected to non-interacting
  leads.  The problem is reformulated in such a way that a
  Landauer-like term appears in the current as well as extra terms
  corresponding to non-equilibrium many-body effects. The interaction
  in the central region renormalizes not only the Green's functions
  but also the coupling at the contacts between the central region and
  the leads, allowing the total current to be further expressed as a
  generalized Landauer-like current formula.  The general expression
  for the dynamical functional that renormalizes the contacts is
  provided.  We analyze in detail under what circumstances
  Landauer-like approaches to the current, i.e. without
  contact renormalization, are valid for interacting electron-electron
  and/or electron-phonon systems.  Numerical NEGF calculations are
  then performed for a model electron-phonon coupled system in
  order to validate our analytical approach.  We show that the
  conductance for the off-resonant transport regime is adequately
  described by Landauer-like approach in the small-bias limit, while
  for the resonant regime the Landauer-like approach results depart from
  the exact results even at small finite bias.  The validity of
  applying a Landauer-like approach to inelastic electron tunneling
  spectroscopy is also studied in detail.
\end{abstract}
 

\maketitle

\section{Introduction}
\label{sec:intro}

Electronic transport through nanoscale systems exhibits many important
new features in comparison with conduction through macroscopic
systems.  In particular local interactions, such as Coulomb
interactions between the electrons and scattering from localized
atomic vibrations, become critically important.  
In crude terms, these effects are more important in nanoscale systems 
as the electronic probability density is concentrated in a small 
region of space; normal screening mechanisms are thus ineffective.  

It is most useful to have a simple expression for the electronic
current or for the conductance of a nanoscale object connected to
terminals.  This is provided in the form an appealing intuitive
physical picture by the Landauer formula\cite{Landauer:1970}, which
describes the current in terms of the transmission coefficients of the
central scattering region and of distribution functions of the
electrons in the terminals.  However, in its original form the
Landauer formula deals only with non-interacting electrons.  This
formalism has been used in conjunction with density-functional theory
(DFT) calculations for realistic nanoscale systems
\cite{Hirose:1994,DiVentra:2000,Taylor:2001,Nardelli:2001,Brandbyge:2002,
  Gutierrez:2002,Frauenheim:2002,Xue:2003,Louis+Palacios:2003,Thygesen:2003,Garcia-suarez:2005}
and has helped tremendously for the qualitative understanding of the
transport properties of such realistic systems. The apparent success
of such approaches relies on the fact that DFT maps the many-electron
interacting system onto an effective non-interacting single-particle
Kohn-Sham Hamiltonian suited for the Landauer formalism for transport.
However when such a mapping becomes questionable for
strongly-interacting electron systems, the original Landauer approach
has been found to be incomplete and unable to properly take into
account the many-body effects \cite{MeirWingreen:1992,Vignale:2009}.

The Landauer formula has been built upon by Meir and
Wingreen \cite{MeirWingreen:1992} to extend this formalism to a central
scattering region containing interacting between particles.
It is then expressed in terms of non-equilibrium Green's functions and self-energies
and in the most general cases it does not bear any formal resemblance with the original
Landauer formula for the current \cite{MeirWingreen:1992,Haug:1996}.
Other generalisation of 
Landauer-like approaches to include interactions and inelastic scattering have
been developed, see for example 
Refs.~[\onlinecite{Imry:2005,Ferretti:2005a,Ferretti:2005b,Vignale:2009}].

It is therefore important to know the domain of validity of
Landauer-like approaches in comparison to exact current calculations
based on non-equilibrium Green's functions for treating electron
transport through an interacting region connected to leads at
different thermodynamical equilibria.
This is the question we address in this paper by following a
two step approach.

First we reformulate Meir and Wingreen's work to once more
express the current as the sum of a Landauer-like expression involving
a transmission coefficient, plus a non-Landauer-like term arising from
the non-equilibrium many-body effects.
We further develop our theoretical framework to show that the interaction between
particles in the central region not only renormalizes the non-equilibrium \GFs but also 
the coupling at the contacts between the central region and the leads. 
We hence obtain a generalized Landauer-like formula for the current in the same 
spirit as in 
Refs.~[\onlinecite{Sergueev:2002,Zhang:2002,Ferretti:2005a,Ferretti:2005b}].
However our result for the dynamical functional that renormalizes the coupling
at the contacts is more general than the ansatz used in previous studies 
(Refs.~[\onlinecite{Ng:1996,Sergueev:2002,Zhang:2002,Ferretti:2005a,Ferretti:2005b}]).
Our result does not imply any constraints on the statistics of the non-equilibrium 
interacting central region.

Secondly, we apply our theoretical framework to study a model system 
in the presence of electron-phonon (electron-vibron) interactions, connected to two 
non-interacting electron leads at non-equilibrium. 
We analyze in detail the validity of Landauer-like approaches to describe the conductance
and the inelastic electron tunneling spectroscopy (IETS) of such a non-equilibrium
many-body interacting system.

The paper is organized into two main sections.  In the first (section
\ref{sec:formalism}) we
develop our formalism to derive our generalized expression for the
current.  The implications of this are discussed in section
\ref{sec:discuss}.  We then apply this formalism to the model system in section
\ref{sec:numer-calc} and show the results of numerical
calculations.

\section{Formalism}
\label{sec:formalism}

\subsection{The model system}
\label{sec:modelsystem}

Following Meir and Wingreen \cite{MeirWingreen:1992}, we consider
a scattering central region (a quantum dot, a molecule, or a nanowire including 
interaction between particles) which
is connected to two (left $L$ and right $R$) leads. These leads are described 
by two non-interacting Fermi seas at their own equilibrium,
characterized by two Fermi distributions $f_L(\omega)$ and 
$f_R(\omega)$.

The Hamiltonian of the system is given by 
\begin{equation}
  \label{eq:MeirWingreenHamiltonian}
 \begin{split}
   \hat{H} =  \sum_{\alpha= L, R} & \varepsilon_{\alpha}
   \hat{c}_{\alpha}^\dagger \hat{c}_{\alpha} + \hat{H}_{\rm
     int}(\{\hat{d}_n^\dagger\} , \{\hat{d}_n \} ; \{\hat{a}_\lambda^\dagger\} ; \{\hat{a}_\lambda \}) \\
   & + \sum_{n,\alpha = L, R} (V_{\alpha,n} \hat{c}_{\alpha}^\dagger
   \hat{d}_n + {\rm H.c.}),
\end{split}
\end{equation}
where the summation indices $\alpha$ run over the left and right leads
($L, R$ respectively) and depending on the choice of representation
over momentum $k$ or lattice site $i$ index, with
$\hat{c}^\dagger_{\alpha} (\hat{c}_{\alpha})$ creating (annihilating)
a non-interacting electron, and $\{\hat{d}_n^\dagger\} ; \{\hat{d}_n
\}$ represent a complete, orthonormal set of states for the
interacting electrons in the central region, and
$\{\hat{a}_\lambda^\dagger\} ; \{\hat{a}_\lambda\}$ represent a set of
bosonic degrees of freedom to which the electrons are coupled in the
central region. These can be more or less extended phonons in a quantum
dot or nanowire, or molecular vibrations (vibrons) in molecules.

There are two main approximations in the Meir and Wingreen approach
to transport. The first is to consider that the interactions are 
localized within the central region.
This leads to specific properties for the self-energies used to
calculate the non-equilibrium many-body \GFs within the basis states
of the central region only.  The self-energies are then obtained as 
the sum of three
contributions: two similar contributions arising from the electronic
coupling of the central region to the left and right leads and the
third arising from the interaction between particles in the
central region.

The second approximation is to consider that the initial correlations
dies out in the long-time limit, and hence a steady state regime can
be reached. It should be noted that a generalization going beyond the
steady state regime has recently been given in Ref.[\onlinecite{Myohanen:2009}].

\subsection{Non-equilibrium Green's functions and Landauer-like formula
for the current}
\label{sec:NEGFandLandauer}

In the steady state, the current $I_L$ flowing at the left contact between 
the left lead and the central 
region is expressed in terms of three non-equilibrium Green's
functions (the retarded $G^r$, advanced $G^a$ and lesser $G^<$ Green's
functions) of the dressed interacting central region \cite{MeirWingreen:1992}.  

Using the identity $G^>-G^< = G^r-G^a$ and the definition of the
leads' self energies $\Sigma_L^<(\omega)= i
f_L(\omega)\Gamma_L(\omega)$ and $\Sigma_L^>(\omega)= -i
(1-f_L(\omega))\Gamma_L(\omega)$, for which we recall that
$f_L(\omega)$ is the Fermi distribution of the non-interacting left
lead
and $\Gamma_L$ is obtained from the imaginary part of the retarded
(advanced) self-energy $\Sigma^{r(a)}_L$ arising from the coupling of
the central region to the left lead, i.e. 
$\Gamma_L(\omega)=\mp 2 \Im m \Sigma^{r/a}_L(\omega)$, 
the current $I_L$ is given by
\cite{Mii:2003,Frederiksen:2004b,
Galperin:2004b,
Mitra:2004,
Pecchia:2004b,
Chen_Z:2005,
Ryndyk:2005,
Sergueev:2005,
Viljas:2005,
Yamamoto:2005,
Cresti:2006,
Vega:2006,
vanLeeuwen:2006,
Egger:2008} 

\begin{equation}
\label{eq:I_MeirWingreen}
\begin{split}
I_L  = \frac{2ie}{\hbar} 
\int \frac{{\rm d}\omega}{2\pi} \
{\rm Tr} & [f_L(\omega)\ \Gamma_L(\omega)\
[G^r(\omega)-G^a(\omega)] \\
&  + \Gamma_L(\omega)\ G^<(\omega)] ,
\end{split}
\end{equation}
where the trace runs over indexes ${n,m}$ appropriately chosen to represent the
electronic states of the central region.

A similar expression can be obtained for the current $I_R$ flowing at
the right contact between the right lead and the central region by 
exchanging the subscript $L\leftrightarrow R$ in
Eq. (\ref{eq:I_MeirWingreen}).  For a current-conserving system, one
then has $I_L=-I_R$.  The famous result of Meir and Wingreen, (Eq.~(6)
in Ref.~\onlinecite{MeirWingreen:1992}), is then obtained by evaluating the
symmetrized current, $I=(I_L-I_R)/2$, to give
\begin{equation}\label{eq:I_MeirWingreen:Eq-2}
  \begin{split}
  I = \frac{ie}{h} \int {\rm d}\omega\  &
    {\rm Tr} 
    \left[ 
      \left( f_L(\omega) \Gamma_L - f_R(\omega) \Gamma_R \right)
      \left( G^r(\omega) - G^a(\omega) \right) 
    \right.
\\
       & \left. +
      \left( \Gamma_L(\omega) - \Gamma_R(\omega) \right) G^<(\omega) \right]  .
  \end{split}
\end{equation}
Now, we can define more explicitly the specific property of the self-energies, namely additivity: 
$\Sigma^x(\omega)=\Sigma^x_L(\omega)+\Sigma^x_R(\omega)+\Sigma^x_{\rm int}(\omega)$
where $x$ is any component $x=r,a,>,<$, and the \SEs are defined within the central region
by $\Sigma^x_\alpha(\omega)$ for the coupling of the central region to the lead $\alpha$,
and by $\Sigma^x_{\rm int}(\omega)$ for the interaction between electrons or between electrons
and phonons/vibrons. As mentioned in the previous section, the many-body interaction self-energy
can be added to the leads' self-energies only because the interactions are localized in the
central region.
Throughout the paper, we will also use a more compact notation for the leads' \SE, i.e. 
$\Sigma^x_{L+R}=\Sigma^x_L+\Sigma^x_R$.

Using the additivity property of the self-energy, and the fact that
$G^<(\omega)=G^r(\omega)\Sigma^<(\omega)G^a(\omega)$ in the steady state regime, 
the symmetrized current $I$ can be re-expressed as follows:
\begin{equation}
\label{eq:I_Landauer+corrections}
\begin{split}
I=\frac{2e}{\hbar} 
\int  & \frac{{\rm d}\omega}{2\pi} \left\lgroup
(f_L-f_R) {\rm Tr}\left[ \Gamma_L G^r \Gamma_R G^a \right] \right. \\ 
& + {\rm Tr}\left[ (f_L\Gamma_L-f_R\Gamma_R) G^r 
\frac{{ i}(\Sigma^>_{\rm int}-\Sigma^<_{\rm int})}{2} G^a \right] \\ 
& + \left. {\rm Tr}\left[ (\Gamma_L-\Gamma_R)\ G^r \frac{{ i}\Sigma^<_{\rm int}}{2} G^a \right] \right\rgroup.
\end{split}
\end{equation}

Introducing the non-equilibrium distribution function matrix for the interaction 
$f^{\rm NE}_{\rm int}$ defined from the interaction \SEs as 
$\Sigma^<_{\rm int} = - f^{\rm NE}_{\rm int}\ ( \Sigma^r_{\rm int} - \Sigma^a_{\rm int})$
(see Appendix \ref{App:distribution_fnc}), 
one can rewrite the symmetrized current $I$ as follows
\begin{equation}
  \label{eq:I_Landauer+corrections_withfNEint}
  \begin{split}
    I = \frac{2e}{\hbar} 
   \int & \frac{{\rm d}\omega}{2\pi} \left\lgroup
    (f_L - f_R) {\rm Tr} \left[ \Gamma_L G^r \Gamma_R G^a \right] \right. \\
   &  + {\rm Tr} \bigg[ \bigg( (f_L - f^{\rm NE}_{\rm int})\Gamma_L - (f_R -
     f^{\rm NE}_{\rm int})\Gamma_R \bigg)  \\
     &  \qquad \qquad \left. \times G^r 
      \frac{{ i}(\Sigma^>_{\rm int} - \Sigma^<_{\rm int})}{2} G^a \bigg] \right\rgroup.
  \end{split}
\end{equation}

The first term in Eq.~(\ref{eq:I_Landauer+corrections}) and (\ref{eq:I_Landauer+corrections_withfNEint}) 
looks like a Landauer-like (LL) expression for the current, 
\begin{equation}
  \label{eq:I_Landauer}
  \begin{split}
    I^{\rm LL} & = \frac{2e}{h} \int {\rm d}\epsilon (f_L(\epsilon)-f_R(\epsilon)) T_{\rm eff}(\epsilon) \\
    & = \frac{2e}{\hbar} 
    \int \frac{{\rm d}\omega}{2\pi} 
    (f_L-f_R)\ {\rm Tr}\left[ \Gamma_L G^r \Gamma_R G^a \right],
  \end{split}
\end{equation}
with an effective transmission
\begin{equation}
  \label{eq:Effective-transmission}
   T_{\rm eff}(\epsilon)= {\rm Tr} \left[ \Gamma_L G^r \Gamma_R G^a \right](\epsilon) 
		  	= {\rm Tr} [t^\dag(\epsilon) t(\epsilon)] ,
\end{equation}
which can be interpreted with the intuitive physical picture, as in the original Landauer formulation 
of electronic transport, in terms of transmission coefficients $t(\varepsilon)$ and propagation 
eigenchannels as defined in Refs.~[\onlinecite{Brandbyge:1997,Paulsson:2007}].
 
The second term in Eq.~(\ref{eq:I_Landauer+corrections_withfNEint}) corresponds to 
\NE corrections due to the interaction. It is expressed in terms of $\Sigma^{<,>}_{\rm int}$ 
and of the different distribution functions. 
This term, not automatically small, cannot be recast in the form of extra transmission coefficients 
as in Refs.~[\onlinecite{Brandbyge:1997,Paulsson:2007}], and already indicates in a way the 
breakdown of the original Landauer formula for the current in the presence of
interaction \cite{MeirWingreen:1992}.

One should note also that even if $I^{\rm LL}$ looks like a 
Landauer formula for the current with an effective transmission
$T_{\rm eff}(\epsilon)$, 
the interaction between particles is already taken into account
in an exact calculation of the Green's functions.
In this sense, $I^{\rm LL}$ is not a conventional Landauer current 
formula for single-particle elastic scattering.
The renormalization of the non-interacting reference system is
included in the retarded and advanced \GFs via the corresponding self-energies:
$G^{r,a}(\omega)=[\ g^{r,a}_0(\omega)^{-1}-\Sigma^{r,a}_{L+R}(\omega) - \Sigma_{\rm int}^{r,a}(\omega) ]^{-1}$.
Depending on the way the interactions are treated, the renormalization of the 
\GFs may even go beyond the quasi-particle description of the interacting system.
In any case, the important point is that $T_{\rm eff}$  already contains part of the 
electron-electron and/or electron-phonon inelastic scattering processes.
 
To complete our theoretical framework, we can make a further formal manipulation of the equations
for the exact current, as given for example by Eq.~(\ref{eq:I_Landauer+corrections_withfNEint}),
and end up with a more compact expression for the current which expresses a clear physical result: 
the interaction renormalizes not only the \GFs ($G^{r,a}$) but also the coupling at the contacts.

To show this, it is more convenient to consider for the moment the current at only one 
contact ($I_L$ for example), 
though one should not forget that in the steady state the current conservation
implies $I_L=-I_R=I$. The compact expression we find for $I_L$ is the following:
\begin{equation}
\label{eq:I_Landauerlike_with_contactrenorm}
I_L=\frac{2e}{\hbar} 
\int \frac{{\rm d}\omega}{2\pi} 
(f_L(\omega)-f_R(\omega))
{\rm Tr} \left[ \Gamma_L G^r \Upsilon_R G^a \right] 
\end{equation}
with the coupling to the right contact $\Upsilon_R$ being renormalized as
\begin{equation}
\label{eq:GammaR_renorm}
\Upsilon_R(\omega)=\Gamma_R(\omega) \Lambda(\omega) ,
\end{equation}
and
\begin{equation}
\label{eq:Lambda}
\Lambda(\omega) =
1+ \Gamma_R^{-1}
\frac{f_L(\omega)-f^{\rm NE}_{\rm int}(\omega)}{f_L(\omega)-f_R(\omega)} \
{ i}(\Sigma^>_{\rm int}-\Sigma^<_{\rm int})(\omega) ,
\end{equation}
where we recall that $f^{\rm NE}_{\rm int}(\omega)$ is the non-equilibrium statistical
distribution for the many-body interactions as defined in Appendix~\ref{App:distribution_fnc}.

Equations (\ref{eq:I_Landauer+corrections_withfNEint})
and (\ref{eq:I_Landauerlike_with_contactrenorm}-\ref{eq:Lambda}) 
(see also Eq.~(\ref{eq:newLambda_2}) in Appendix~\ref{App:Ng_ansatz}) represent the principal 
formal results of this paper. 
They imply that for an interacting central region, one can always express the current 
in an generalized Landauer-like formula in which not only the retarded and advanced \GFs 
are renormalized by the interaction but also the coupling at the contacts,
as similarly found in Refs.~[\onlinecite{Ng:1996, Sergueev:2002, Zhang:2002, Ferretti:2005a, Ferretti:2005b}].
This generalized formula needs to be contrasted with the more conventional Landauer-like formula 
Eq.~(\ref{eq:I_Landauer}) in which the contacts of the central region with the leads are not 
renormalized by the interaction. 

Our expressions Eq.~(\ref{eq:I_Landauerlike_with_contactrenorm}-\ref{eq:Lambda}) are valid for 
any kind of interaction localized in the central region and generalize the results of the previous 
studies (Refs.~[\onlinecite{Ng:1996, Sergueev:2002,Zhang:2002, Ferretti:2005a, Ferretti:2005b}]) 
because they do not imply any restrictions to the non-equilibrium statistics of the many-body 
interacting central region as we explain in detail in Appendix~\ref{App:Ng_ansatz}.

Finally, one recovers the more conventional Landauer-like formula (with no
correction factors or equivalently with no renormalization of the contact couplings)
when the quantity $\left(\Sigma^>_{\rm int}(\omega)-\Sigma^<_{\rm int}(\omega)\right)$ vanishes,
as can be clearly seen from Eqs.~(\ref{eq:I_Landauer+corrections_withfNEint}) and 
Eqs.~(\ref{eq:I_Landauerlike_with_contactrenorm}-\ref{eq:Lambda}).
In the next section we discuss in detail the conditions for which this can happen.
Note that the condition $(\Sigma^>_{\rm int}(\omega)-\Sigma^<_{\rm int}(\omega))=0$ does not
necessarily imply that $G^{r,a}(\omega)=0$ as well. 
Hence, even when the transport is well described by a Landauer-like formula  
($(\Sigma^>_{\rm int}(\omega)-\Sigma^<_{\rm int}(\omega))=0$), normalization effects still occur
and the transport is dominated by single-quasiparticle scattering.

\subsection{Discussion}
\label{sec:discuss}

Clearly whenever $(\Sigma^>_{\rm int}-\Sigma^<_{\rm int})(\omega)=0$,
there is no renormalization at the contact, and the current is simply
given by $I^{\rm LL}$.  This may happen in two cases: either
$\Sigma^{<,>}_{\rm int}=0$ for all $\omega$ or only within finite
range(s) of $\omega$.  In the latter case, the relevant range of
$\omega$ for which $(\Sigma^>_{\rm int}-\Sigma^<_{\rm int})=0$ should
be included within the bias window defined by the two Fermi levels
$\mu_L$ and $\mu_R$ at non equilibrium.

In order to understand how and why the quantity $\Sigma^>_{\rm int}-\Sigma^<_{\rm int}$
can vanish for an interacting system, let us first come back to the definition of the 
lesser and greater self-energies. These are specific components (projections onto the real time
axis) of the more general \SE $\Sigma_{\rm int}(\tau,\tau')$ with times $\tau,\tau'$ defined on
the Keldysh time-loop contour \cite{Keldysh:1965,Craig:1968,vanLeeuwen:2006,Rammer:2007}.
Within the Keldysh approach, the lesser $<$ (greater $>$) components of $\Sigma_{\rm int}(\tau,\tau')$ 
imply that the times $\tau/\tau'$ are located on the forward/backward (backward/forward respectively)
time-ordered branch. 
$\Sigma^{<,>}(\omega)$ is simply the Fourier transform of $\Sigma^{<,>}_{\rm int}(t,t')$
in the limit of the steady-state regime where any quantity depends only on the time
difference $X(t,t')=X(t-t')$.

First let us examine the first case: why would a self-energy have no 
lesser or greater components?
For the so-called irregular \SEs \cite{Wagner:1991}, we have the condition
$\Sigma(\tau,\tau')=\hat\Sigma(\tau)\ \delta(\tau-\tau')$. 
The self-energies for the interaction are instantaneous (local) in time.
Hence they cannot have lesser or greater components, since the times have to be 
on the same time-loop branch.
This condition of locality in time corresponds to two classes of physical effets. 
First when the \SEs describe one-particle potentials due to electron-electron or electron-phonon 
interaction, in other words it corresponds to the Hartree-Fock approximation for electron-electron 
interaction and to only the Hartree-like approximation for electron-phonon interaction. 
And second when the \SEs correspond to the so-called initial correlations which contain all 
contributions singular in time (see for example Refs. [\onlinecite{Semkat:1999,Semkat:2000}]).

There is also another class of problems for which there are no lesser or greater
components of the self-energy. 
It is when the exchange and correlation effects for interacting 
electron systems are represented by an effective potential $v_{xc}(r,t)=\delta A_{xc}[n]/\delta n(r,t)$
being obtained from an exchange-correlation action functional $A_{xc}[n]$ of the electron 
density $n(r,t)$. To this potential will correspond an effective self-energy that is local 
in both space and, more importantly, in time \cite{Stefanucci:2004a}; 
hence with no lesser and greater components for a generalization onto the Keldysh contour.

In effect, any method which maps an interacting electron/phonon system onto an effective 
one-particle (quasi-particle) scheme as for example in density-functional based technique 
(DFT, TDDFT) or other mean-field approaches, will end up with no lesser and greater components 
for the corresponding self-energy describing the interaction. 
Hence a Landauer-like approach to the transport is entirely appropriate for such methods
\cite{Brandbyge:2002,Louis+Palacios:2003}.
However the mapping onto a one-particle scheme may not describe well strongly correlated 
electronic systems, $A_{xc}[n]$ being amenable to approximation, or is simply not possible 
in the general case of electron-phonon interaction.

Now, let us turn to the second case: the interaction spectral density 
$\Im m \Sigma^r_{\rm int}=i(\Sigma^>_{\rm int}-\Sigma^<_{\rm int})/2$ vanishes for one
or more (connected) ranges of $\omega$ values. 
If this gap in $\Im m \Sigma^r_{\rm int}$ is enclosed within the bias window then again, 
the current $I$ will be determined only by the Landauer-like term $I^{\rm LL}$.

As will be shown in detail below from numerical calculations for a model system, such a gap in
$\Im m \Sigma^r_{\rm int}$ may exist in special cases of electron-phonon interaction. The
gap in $\Im m \Sigma^r_{\rm int}$ is then usually located around the Fermi level at equilibrium,
and around the Fermi levels for the non-equilibrium cases at low applied bias only.
These cases correspond to the regime studied by Imry {\it et al.} \cite{Imry:2005} who 
derived a Landauer-like inelastic transmission for interacting electron-phonon systems and argued 
that the Landauer picture is still valid in the presence of interaction as long as multi-particle 
processes can be neglected.

For electron-electron interactions, the situation is somewhat
different.  Even if collective excitations like plasmons present some
qualitative bosonic analogy to phonons (there is a peak in the \SE
around the plasmon energy---as for the e-ph \SE---and no much
interaction spectral density elsewhere), there is always however a
non-zero contribution to the \SE coming from the continuum of
electron-hole excitations.

The main difference from the electron-phonon interaction is that the phonon 
frequency $\omega_0$ imposes a restricted energy scale on the interaction, while for 
electron-electron interaction all energy scales are available, making the corresponding
interaction self-energy not vanishing, except in a infinitely small energy window around 
the Fermi energy at equilibrium.  
So in principle, Landauer-like approaches for interacting electron systems are not valid for interactions
treated beyond the mean-field/density-functional-based approximations \cite{Vignale:2009}. 

Finally, one should note that the second term in Eq.~(\ref{eq:I_Landauer+corrections_withfNEint}) 
for the current, or the second term in the renormalization function $\Lambda(\omega)$ in 
Eq.~(\ref{eq:Lambda}) involves the quantity $(\Sigma^>_{\rm int}-\Sigma^<_{\rm int})$ which is, 
in a series expansion of the interaction, proportional to the powers of the coupling constant(s) 
characterizing the interaction.
In the limit of weak interactions, these terms represent small corrections to the Landauer-like
current expression, and also give small contributions in the renormalization of both the \GFs and
the coupling at the contacts. Hence in the limit of weak interactions, conventional Landauer-like 
approaches could be confidently used and corrected by using perturbation theory for the 
interaction \cite{Lorente:2000,Chen:2004,Chen_Y:2005}.

Now we turn to presenting numerical calculations for a model system
including electron-phonon interactions to illustrate our previous
analysis. We compare results obtained from the exact current
expression Eq.~(\ref{eq:I_Landauer+corrections_withfNEint}) or
(\ref{eq:I_Landauerlike_with_contactrenorm}) with the current derived
from the Landauer-like formula Eq.~(\ref{eq:I_Landauer}), and by using
different levels of approximations for the Green's functions.

\section{Application for interacting electron-phonon model systems}
\label{sec:numer-calc}

In this section, we study in detail the validity of Landauer-like
approaches for an interacting model system connected to two
non-interacting electron reservoirs at non-equilibrium.  We
concentrate on a model of electron-phonon interaction with the
simplest version of the Hamiltonian (Eq.~(\ref{eq:MeirWingreenHamiltonian}))
for the central part: a single electron level
coupled to a single vibration mode---the single-site single-mode
(SSSM) model \cite{Dash:2010-I}, which has also been considered in
previous studies \cite{Mii:2003,Frederiksen:2004b, Galperin:2004b,
  Mitra:2004, Pecchia:2004b, Chen_Z:2005, Ryndyk:2005, Sergueev:2005,
  Viljas:2005, Yamamoto:2005, Cresti:2006, Vega:2006,Egger:2008}.  We
also briefly describe below how to calculate the non-equilibrium
Green's functions (NEGF) from this model Hamiltonian; the full
theoretical details can be found elsewhere \cite{Dash:2010-I}.

We then apply our NEGF technique to the calculations of the transport
properties of the junction around equilibrium and out of equilibrium,
and thus for different transport regimes. We analyze in detail if and
when Landauer-like approaches can provide a good description of the
transport properties in comparison to an exact calculation.

\subsection{Model Hamiltonian for electron-phonon coupling}
\label{sec:eph-Hamiltonian}

The Hamiltonian for the central region for the SSSM model is then
given by
\begin{equation}
  \label{eq:H_central}
  H_C = \varepsilon_0 d^\dagger d + \hbar \omega_0 a^\dagger a +
  \gamma_0 (a^\dagger + a) d^\dagger d,
\end{equation}
where one electronic level $\varepsilon_0$ and one vibration mode 
of energy $\omega_0$ are coupled together via the coupling constant $\gamma_0$.  

Furthermore we choose a simple model for the structure of the leads, which 
provides analytical results for the corresponding surface Green's
functions, but in principle there is no particular restriction to be applied
to the model or dimensionality of the leads.
So in the following, the left $L$ and right $R$ leads are described by two 
non-interacting one-dimensional semi-infinite tight-binding chains:
\begin{equation}
  \label{eq:H_LRleads}
  \begin{split}
  H_L = \sum_{i=-1}^{-\infty} \varepsilon_L c^\dagger_i c_i + \beta_L
  \left( c^\dagger_i c_{i-1} + {\rm c.c.} \right) \\
  H_R = \sum_{i=+1}^{+\infty} \varepsilon_R c^\dagger_i c_i + \beta_R
  \left( c^\dagger_i c_{i-1} + {\rm c.c.} \right) .
  \end{split}
\end{equation}
This model provides us with analytical expressions for the matrix elements of
the leads' Green's functions at the terminal sites \cite{English:1998}:
\begin{equation}
  \label{eq:g0r_leads}
  g^r_{0\alpha}(\omega) = e^{ ik_\alpha(\omega) } /\beta_\alpha,
\end{equation}
with $\omega=\varepsilon_\alpha+2\beta_\alpha\cos k_\alpha(\omega)$,
giving rise to semi-elliptic density of states of the terminal lead
sites connected to the central region.  

The expression for the
coupling of the central region to the $L$ and $R$ leads is then given by
\begin{equation}
  \label{eq:coupling-pot}
  V_{LC} + V_{CR} = \sum_{\alpha = L, R} t_{0\alpha} (c^\dagger_\alpha
  d + d^\dagger c_\alpha),
\end{equation}
with hopping integrals $t_{0\alpha}$ and $c_{\alpha=L}=c_{i=-1}$,
$c_{\alpha=R}=c_{i=+1}$.

\subsection{Non-equilibrium Green's functions for electron-phonon
coupled system}
\label{sec:NEGF}

We use a non-equilibrium Green's function (NEGF) technique to
calculate the properties of the system in a similar manner to previous studies 
\cite{Mii:2003,Frederiksen:2004b,
Galperin:2004b,
Mitra:2004,
Pecchia:2004b,
Chen_Z:2005,
Ryndyk:2005,
Sergueev:2005,
Viljas:2005,
Yamamoto:2005,
Cresti:2006,
Vega:2006,Egger:2008}.
The details of our NEGF
calculations are described in detail in Ref.~[\onlinecite{Dash:2010-I}], 
but we briefly summarize our application of them here.

The Green's functions are calculated via Dyson-like equations for
the retarded and advanced Green's functions $G^{r,a}(\omega)$:
\begin{equation}
  \label{eq:Greens-retarded-advanced}
  G^{r,a}(\omega) = g^{r,a}_C(\omega) + g^{r,a}_C(\omega) \Sigma^{r,a}(\omega) G^{r,a}(\omega),
\end{equation}
where $g^{r,a,}_C$ is the non-interacting Green's function for the
isolated central region.

For the greater $G^>(\omega)$ and lesser $G^<(\omega)$ Green's functions, we use a
quantum kinetic equation of the form
\begin{equation}
  \label{eq:Greens-greater-lesser}
  G^{>,<} = (1 + G^r \Sigma^r) g^{>,<}_C (1 + \Sigma^a G^a) + G^r
  \Sigma^{>,<} G^a.
\end{equation}
Here $\Sigma^x(\omega), (x = r, a, >, <)$ is a total self-energy consisting of
a sum of the self-energies from the constituent parts of the system:
\begin{equation}
  \label{eq:total-self-energy}
  \Sigma^x(\omega) = \Sigma^x_L(\omega) + \Sigma^x_R(\omega) + \Sigma^x_{\text{int}}(\omega).
\end{equation}
The leads' self-energies $\Sigma^x_{L+R}(\omega)=\Sigma^x_L(\omega)+\Sigma^x_R(\omega)$ arising from the
non-interacting leads $\alpha = L,R$ are given by
\begin{equation}
  \label{eq:Lead-self-energies}
  \begin{split}
  \Sigma^r_\alpha(\omega) & = t_{0\alpha}^2 g^r_{0\alpha}(\omega) = (\Sigma^a_\alpha(\omega))^*, \\
  \Sigma^<_\alpha(\omega) & = -2i\ \Im m[\Sigma^r_\alpha(\omega)] f_\alpha(\omega), \\
  \Sigma^>_\alpha(\omega) & = -2i\ \Im m[\Sigma^r_\alpha(\omega)] (f_\alpha(\omega)-1), \\
  \end{split}
\end{equation}
where $ g^r_{0\alpha}$ is given by Eq.~(\ref{eq:g0r_leads}) and 
$f_\alpha$ is the Fermi-Dirac distribution for lead $\alpha$,
with Fermi level $\mu_\alpha=\mu^{\rm eq}+\eta_\alpha eV$ and 
temperature $T_\alpha$.
At equilibrium, the whole system has a single and well defined Fermi level $\mu^{\rm eq}$.
Out of equilibrium, a finite bias is applied throughout the junctions. Within our model
Hamiltonian, the fraction of electrostatic potential drop at the left contact is
$\eta_L=\pm \eta_V$ and $\eta_R=\mp (1-\eta_V)$ at the right contact \cite{Datta:1997}, 
hence $\eta_L-\eta_R=eV$ is indeed the applied bias, and $\eta_V \in [0,1]$.

The self-energy for the interaction in the central region,
$\Sigma_\text{int}(\omega)$, is obtained from a non-equilibrium
many-body perturbation expansion \cite{{Dash:2010-I}} of the
electron-phonon coupling term in the Hamiltonian,
Eq.~(\ref{eq:H_central}).  As for a conventional many-body perturbation
expansion, the self-energy is associated with a series of Feynman
diagrams for the interaction.  In the current work, we consider only
the lowest order diagrams, i.e. the Born approximation (BA) or
equivalently the Hartree-Fock approximation \cite{Mii:2003,
  Frederiksen:2004b, Galperin:2004, Mitra:2004, Pecchia:2004,
  Chen_Z:2005, Ryndyk:2005, Sergueev:2005, Viljas:2005, Yamamoto:2005,
  Cresti:2006, Vega:2006, Egger:2008}.
The exact expressions for $\Sigma^x_\text{int}, (x = r, a, >, <)$ at the Hartree-Fock level 
and beyond are given in Ref.~[\onlinecite{{Dash:2010-I}}] and we do not reproduce them 
here.

\subsection{Numerical results}
\label{sec:num_res}

We divide the calculations into two different types of transport regimes. The first of these is 
when either $\varepsilon_0 \ll \mu^{\rm eq}$ or  $\varepsilon_0 \gg
\mu^{\rm eq}$, known as the off-resonant regime. 
It corresponds to a poorly conducting junction (i.e. semiconductor or insulator-like behavior) dominated 
by strong tunneling at low bias.
The second transport regime is when $\varepsilon_0 \sim \mu^{\rm eq}$ and known as the resonant 
transport regime. This regime corresponds to a good, metallic-like, conducting junction.

We will see below that depending on the nature of the transport
regime, the Landauer-like approaches may be sufficient, under certain
conditions, to describe the conductance $G(V)=dI/dV$ or the
inelastic electronic tunneling spectroscopy (IETS) properties of the
junctions.

\subsubsection{The self-energy quantity $(\Sigma^>_{\rm int}-\Sigma^<_{\rm int})$}
\label{sec:SEquantity}

\begin{figure}		
\begin{center}
\includegraphics[width=\columnwidth]{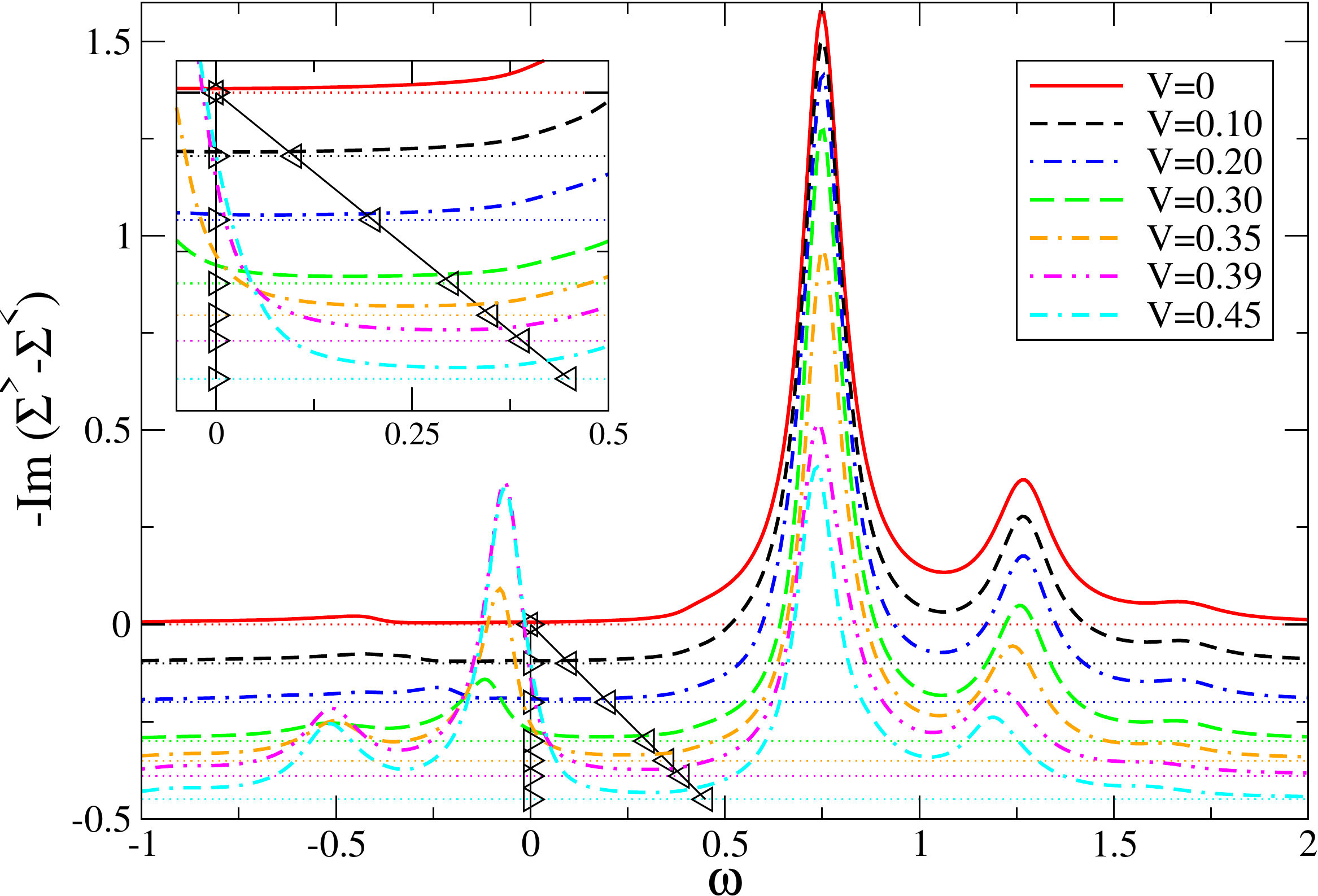}
\end{center}
\caption{The $(\Sigma^>_{\rm int}-\Sigma^<_{\rm int})(\omega)$
  quantity for the SSSM model within NEGF-SCBA for off-resonant
  transport regime and intermediate electron-phonon coupling strength
  $\gamma_0/\omega_0=0.65$.  The curves are obtained for different
  applied bias $eV=\mu_L-\mu_R$, and are offset vertically for
  clarity.  The applied bias is also shown and given by the chemical
  potentials of the left $\mu_L$ (left-pointing arrows) and right $\mu_R$
  (right-pointing arrows) leads respectively.  The other parameters
  are $\varepsilon_0=+0.5, \omega_0=0.4, \gamma_0=0.26, t_{0L,R}=0.2,
  T_{L,R}=0.011, \eta=0.025, \eta_V=1$.  At zero and low bias (here
  $V\lesssim 0.25$) the interaction \SEs difference $(\Sigma^>_{\rm
    int}-\Sigma^<_{\rm int})$ is zero within the bias window, and
  hence the current is given by a Landauer-like formula.  }
\label{fig:SCBAsigma_int_gMl_offres}
\end{figure}

As discussed in section \ref{sec:discuss}, the quantity
$(\Sigma^>_{\rm int}-\Sigma^<_{\rm int})$ plays the key role in
determining whether or not the Landauer-like approaches are valid.
This quantity is plotted in Figure \ref{fig:SCBAsigma_int_gMl_offres}
for the off-resonant transport regime and for intermediate
electron-phonon coupling strength. Calculations were performed
self-consistently using the lowest order electron-phonon diagrams,
i.e. within the conventional self-consistent Born approximation (SCBA)
\cite{Dash:2010-I}.  By definition $\Sigma^>_{\rm int}-\Sigma^<_{\rm
  int}=\Sigma^r_{\rm int}-\Sigma^a_{\rm int}$ is a purely imaginary
function for conserving approximations
\cite{Baym:1962,Kadanoff:1962,Strinati:1988,KwongBonitz:2000}.  
It presents features (peaks) corresponding to the excitations of the
system. The features obtained for non-equilibrium conditions, especially 
when real excitations can be created in the system (applied bias 
$V \ge \omega_0$), are strongly different than the features obtained
for equilibrium (no applied bias $V=0$).

At zero and low bias ($V \lesssim 0.6\ \omega_0$ for the set of
parameters used in Figure \ref{fig:SCBAsigma_int_gMl_offres}), the
difference between the interaction \SEs $(\Sigma^>_{\rm int}-\Sigma^<_{\rm int})$
is virtually zero within the bias window $[\mu_R,\mu_L]$.  This means that the
current in Eq.~(\ref{eq:I_Landauer+corrections_withfNEint}) is effectively given
only by the first term $I^{\rm LL}$, and hence Landauer-like approaches
are sufficient to describe the transport properties of the system for
such biases.

\subsubsection{Conductance and inelastic electron tunneling spectroscopy (IETS)}
\label{sec:conductance+IETS}

\begin{figure} 		
\includegraphics[width=\columnwidth]{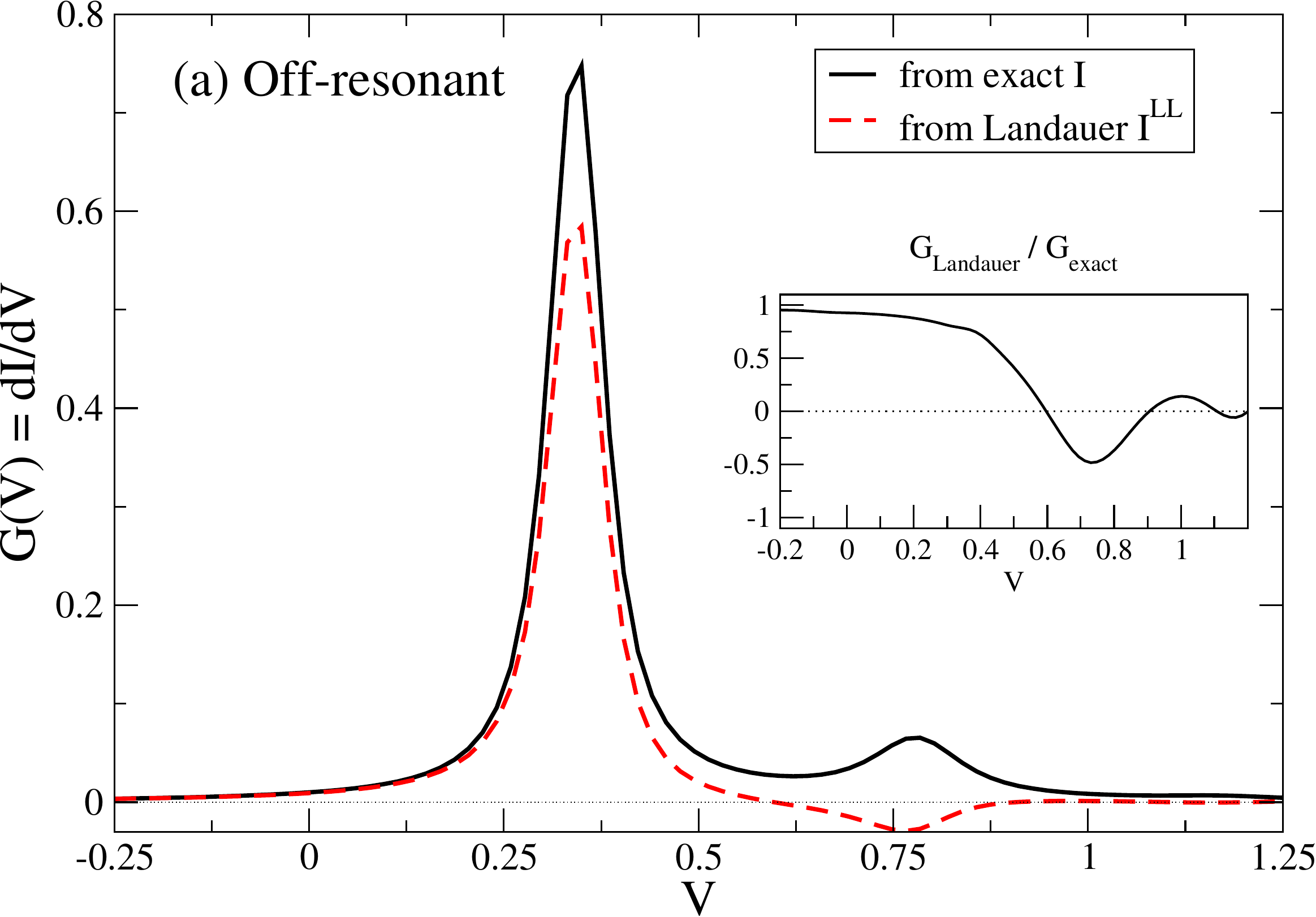}
\includegraphics[width=\columnwidth]{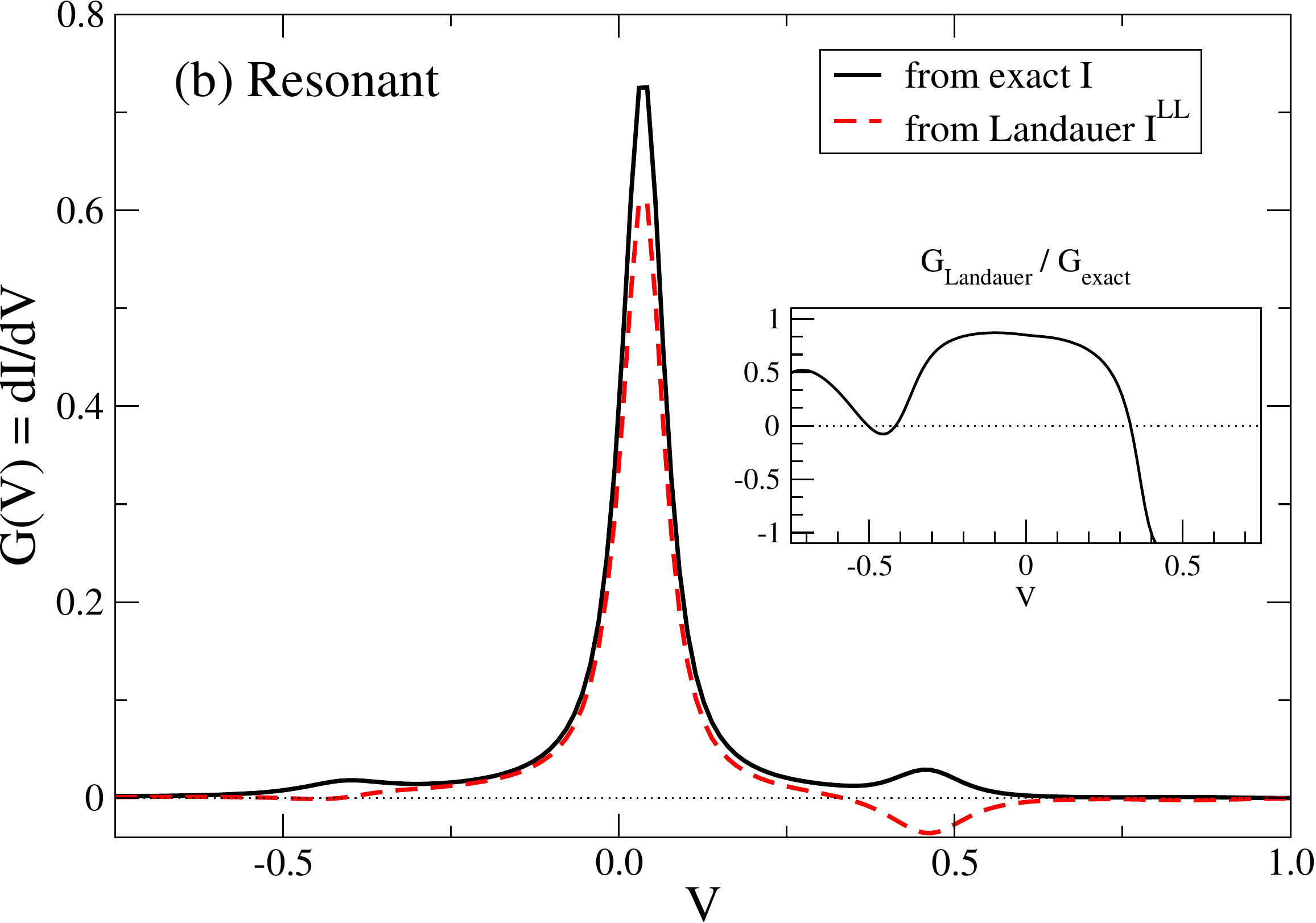}
\caption{Dynamical conductance $dI/dV$ from the exact expression for the current $I(V)$ 
and the corresponding Landauer-like current $I^{\rm LL}(V)$.
\GFs calculations are performed self-consistently for the electron-vibron coupling 
within the Born approximation and for intermediate coupling
$\gamma_0/\omega_0=0.65$.
In the off-resonant case (upper panel, $\varepsilon_0=+0.5$), there is a good agreement 
between the two conductances for the linear regime and for low biases, i.e. when 
$(\Sigma^>_{\rm int}-\Sigma^<_{\rm int})(\omega)\sim0$ as shown on
figure \ref{fig:SCBAsigma_int_gMl_offres}. 
Hence Landauer-like approaches are valid in the low bias regime for off-resonant transport.
For larger biases ($V>0.5$), $dI^{\rm LL}/dV$ gives unphysical negative conductance values.
In the quasi-resonant case (lower panel, $\varepsilon_0=+0.2$),
only the linear conductance is well reproduced by $I^{\rm LL}$. Large deviations of 
$dI^{\rm LL}/dV$ from the exact result occur at small applied bias. Hence the
Landauer-like formula for non-equilibrium current is not valid for the quasi-resonant case.
The other parameters for the calculations are the same as used in 
figure \ref{fig:SCBAsigma_int_gMl_offres}: 
    $\omega_0=0.4, \gamma_0=0.26, t_{0L,R}=0.2, \eta_V=1$.
The insets show the ratio of the conductances calculated
with the exact and the corresponding Landauer-like current.}
\label{fig:IV_and_IVLandauer_offres}
\end{figure}

In order to analyze in detail the different contributions to the conductance and
the conditions for which the Landauer-like approaches can be valid, we have 
performed calculations for the current within different levels of approximation.
In the following, we consider four different kinds of approximation:
firstly, in the absence of interaction, the current is obtained from the 
non-interacting \GFs 
$G_0^{r,a}(\omega)=[ g^{r,a}_0(\omega)^{-1}-\Sigma^{r,a}_{L+R}(\omega)]^{-1}$
and corresponds to the original Landauer formulation $I^{\rm LL}[G_0]$ calculated
with Eq.~(\ref{eq:I_Landauer}). 

Secondly, the Landauer current $I^{\rm LL}[G_0]$ can be corrected to include the 
interaction effects at the lowest order of the coupling parameters, as in perturbation 
theory extended to non-equilibrium conditions. 
This is done by calculating the current in Eq.~(\ref{eq:I_Landauer+corrections}) using 
only the non-interacting \GFs $G_0^{r,a}$
in the first term of Eq.~(\ref{eq:I_Landauer+corrections}) and in the evaluation
of the interaction \SEs $\Sigma_{\rm int}^{<,>}[G_0]$; we denote this current by 
$I_{\rm perturb}=I^{\rm LL}[G_0]+\Delta I[\Sigma_{\rm int}[G_0]]$.

Then the last two kinds of approximations include full renormalization effects in the 
Green's functions.
The first of these corresponds to a non self-consistent BA calculation using the \SEs 
$\Sigma_{\rm int}[G_0]$ to renormalize the \GFs as follows:
$G^{r,a}_{BA}(\omega)=[ G^{r,a}_0(\omega)^{-1} - \Sigma_{\rm int}^{r,a}[G_0(\omega)] ]^{-1}$.
These \GFs are then used to calculate the current 
$I[G_{BA}]=I^{\rm LL}[G_{BA}]+\Delta I[\Sigma_{\rm int}[G_0]]$. 
Finally, the last approximation, correspond to a fully self-consistent renormalization SCBA calculation
performed as described in Ref.~[\onlinecite{Dash:2010-I}], from which we obtain the exact current 
$I[G_{SCBA}]=I^{\rm LL}[G_{SCBA}]+\Delta I[\Sigma_{\rm int}[G_{SCBA}]]$.

The dynamical conductance $G(V)$ is obtained as usual from the first
derivative of the current versus the applied bias $G(V)=dI/dV$.
Typical examples for both the off-resonant and resonant transport
regimes are shown in Figure \ref{fig:IV_and_IVLandauer_offres} where
we compare the conductance obtained from the exact current with the
Landauer-like current $I^{\rm LL}(V)$ using full SCBA
calculations.

For the off-resonant transport regime
(Fig.~\ref{fig:IV_and_IVLandauer_offres}(a)), which is dominated by
strong tunneling at low biases, there is a good agreement between the
exact conductance and the Landauer-like conductance for both the
linear regime and the non-linear regime at low biases, i.e. when
$(\Sigma^>_{\rm int}-\Sigma^<_{\rm int})(\omega)\sim0$ (as shown on
Figure \ref{fig:SCBAsigma_int_gMl_offres}).  Hence in this case, 
Landauer-like approaches are valid to describe the tunneling
transport properties in the low-bias regime for off-resonant
transport.  However, strong discrepancies between the two conductances
occur for biases around the first renormalized electronic resonance
$V\sim 0.3$, even before real excitations of the phonon mode are
available. For even larger biases, ($V \gtrsim 0.5$), $dI^{\rm LL}/dV$ gives
unphysical negative conductance values.

For the quasi-resonant transport regime
(Fig.~\ref{fig:IV_and_IVLandauer_offres}(b)), where the renormalized
electronic resonance $\epsilon_0-\gamma_0^2/\omega_0\sim 0$ is close
to the Fermi level at equilibrium, only the linear conductance is well
reproduced by $I^{\rm LL}$. This is essentially due to the fact that
within conservation approximations for the self-energies, the linear
conductance is not renormalized by the interaction \cite{Oguri:1997,Cornaglia:2004}, 
as we have also shown in detail in the appendix of Ref.~[\onlinecite{Dash:2010-I}].
However large deviations of the Landauer conductance from the exact
result occur quickly at small applied bias. Hence Landauer-like
approaches for non-equilibrium current are not valid for the
quasi-resonant case, and probably not as well for the resonant
case.

\begin{figure} 
\begin{center}
\includegraphics[width=\columnwidth]{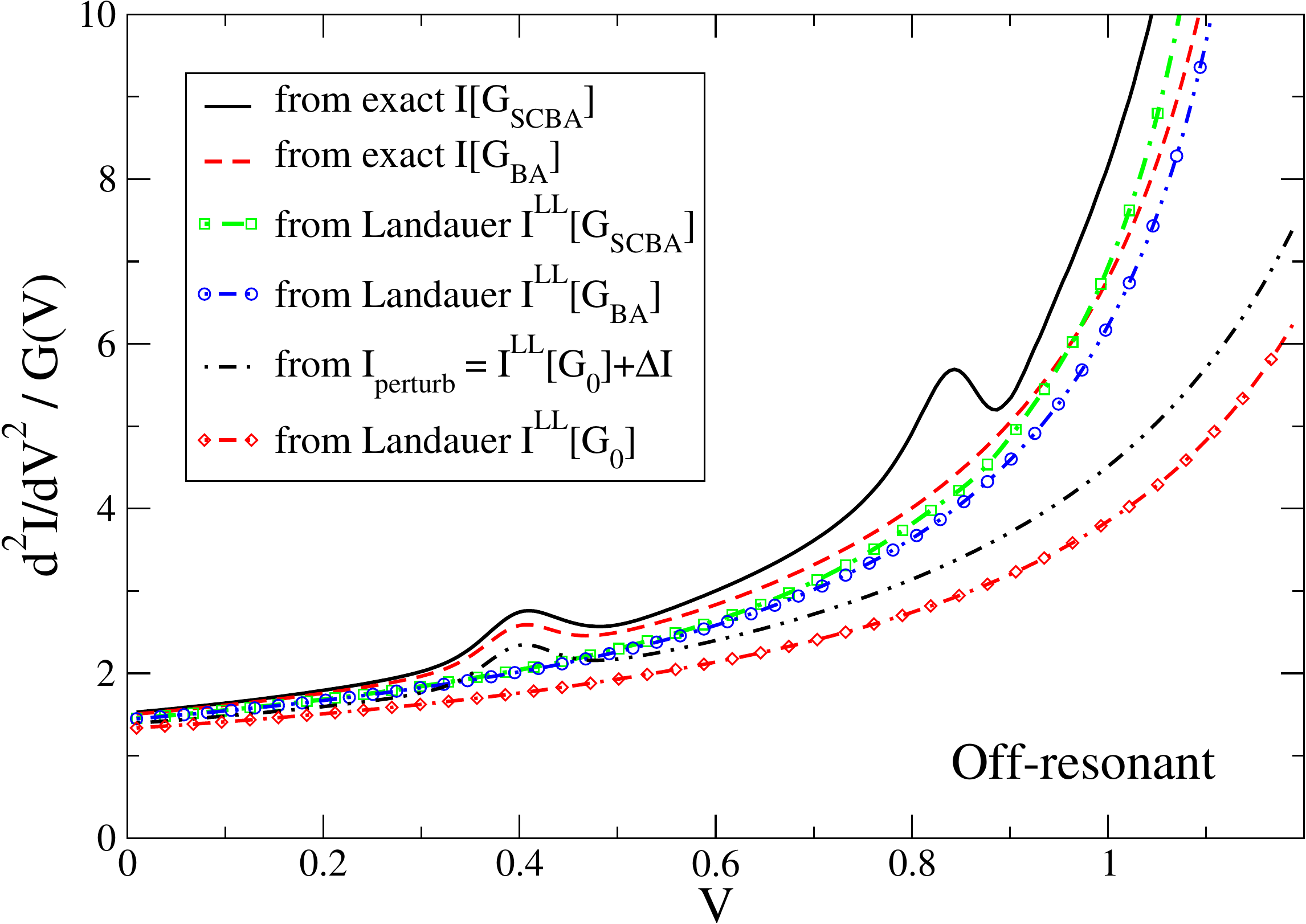}
\end{center}
\caption{IETS signal $d^2I/dV^2$, normalized by dynamical conductance
  $G(V)=dI/dV$, obtained from the exact expression for the current $I(V)$ and the
  corresponding Landauer-like term $I^{\rm LL}(V)$.  Calculations are
  performed for the off-resonant transport regime and the strong
  coupling limit $\gamma_0/\omega_0=0.875$ using different level of
  approximations for the Green's functions, as explained in the legend (see main
  text for detail).  The inelastic vibron excitations are present
  in the IETS derived from the exact $I(V)$ and are located at integer
  multiples of the vibron energy $\omega_0$. They give a positive
  contribution to the baseline, as expected for opaque junctions
  (insulator/semiconductor-like behavior).  However this signal is
  absent in the IETS derived from the Landauer-like term $I^{\rm
    LL}(V)$. This implies that even with normalized \GFs the
  Landauer-approaches are not able to correctly reproduce  the
  inelastic features in IETS for the off-resonant regime.  The
  parameters for the calculations are $\varepsilon_0=+1.5,
  \omega_0=0.4, \gamma_0=0.35, t_{0L,R}=0.11, T_{L,R}=0.011,
  \eta=0.02, \eta_V=1$.  }
\label{fig:IETS_I_and_ILandauer_offres}
\end{figure}

We now turn to the inelastic electron tunneling spectrum (IETS) which
gives information about the selective excitation of the system.  The
IETS is usually obtained from the second derivative of the current
with respect to the applied bias $d^2I/dV^2$.  The IETS curves present
features, peaks or dips \cite{Galperin:2004b}, at biases corresponding
to the energy of a specific excitation, in our case to the energy of
one or several excitations of the vibration mode $n \omega$.  Being
the derivative of the conductance, the IETS curves also present
features at biases corresponding to peaks in the conductance (see for
example Figure \ref{fig:IV_and_IVLandauer_offres}).

\begin{figure}
\begin{center}
\includegraphics[width=\columnwidth]{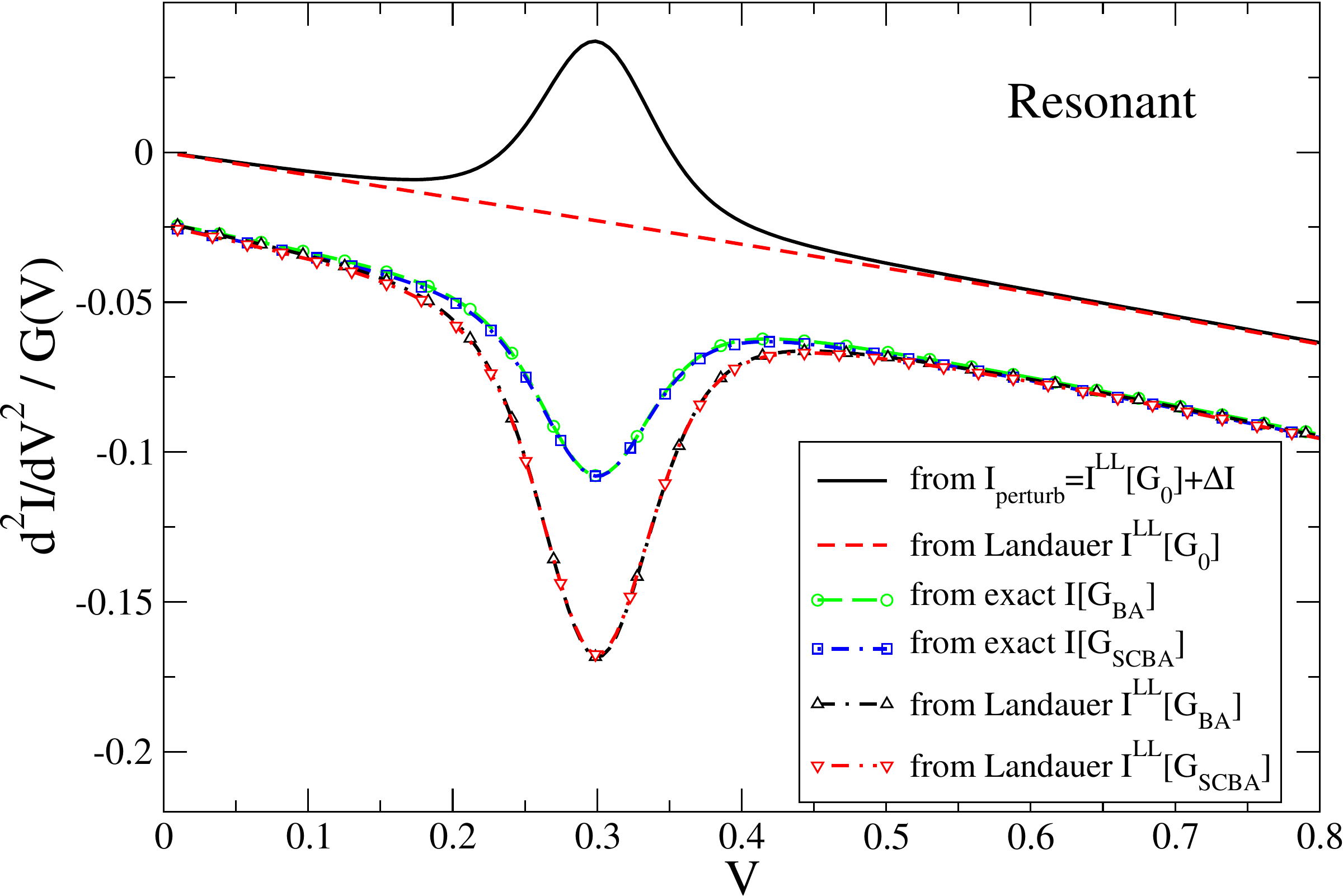}
\end{center}
\caption{IETS signal $d^2I/dV^2$, normalized by $G(V)$, obtained from the exact 
current $I(V)$ and from the corresponding Landauer-like $I^{\rm LL}(V)$ term.
Calculations are for the resonant transport regime, intermediate electron-vibron
coupling $\gamma_0/\omega_0=0.65$ and strong coupling to the leads.
The different approximations used to calculate the \GFs are shown in the legend
(see main text for detail).
The inelastic vibron excitation is present in the IETS signal derived from the
exact $I(V)$ and is located around the vibron energy $\omega_0$. It corresponds 
to a negative contribution to the baseline, as expected
for mostly transparent junctions (metallic-like behavior).
Interestingly, this feature is also present in the IETS derived 
from $I_{\rm LL}(V)$ in contrast to what is obtained for the off-resonant case.
Hence, for resonant transport, it seems that Landauer-like approach can reproduce
the inelastic IETS features at $V=\omega_0$.
The parameters for the calculations are
    $\varepsilon_0=0, \omega_0=0.3, \gamma_0=0.195, t_{0L,R}=1.50,
    T_{L,R}=0.011,
    \eta=0.025, \eta_V=1$.
}
\label{fig:IETS_I_and_ILandauer_res}
\end{figure}

We have found that in order to get a better aspect ratio for the IETS
features corresponding to phonon excitations, it is more convenient
to normalize the IETS curves by the dynamical conductance, i.e.
$[d^2I/dV^2]/[dI/dV]=d/dV \ln G(V)$.  Typical examples for both the
off-resonant and resonant transport regimes are shown in Figures
\ref{fig:IETS_I_and_ILandauer_offres} and
\ref{fig:IETS_I_and_ILandauer_res} respectively.

We compare the IETS signals obtained from the four approximations used
to calculate the current: $I_{\rm Landauer}$ for the Landauer current of
the non-interacting system, $I_{\rm perturb}$ for the Landauer current
corrected by first order perturbation theory for the electron-phonon
coupling, and full renormalization within a non-self-consistent and
self-consistent scheme $I[G_{BA}]$ and $I[G_{SCBA}]$ from which the
corresponding Landauer-like contribution can be extracted.

We first comment on the results obtained for the off-resonant case,
Figure \ref{fig:IETS_I_and_ILandauer_offres}(a).  As expected, the
original Landauer approach does not provide any feature at the bias
$V=\omega_0$ since there is no interaction. 
The IETS signal obtained from the fully self-consistent calculations
shows however two features (peaks in the case of strong tunneling regime 
at low bias \cite{Galperin:2004,Galperin:2004b}) at biases $V=\omega_0$ and $V=2 \omega_0$.
They correspond to inelastic processes involving the excitation of 
one and two vibration modes respectively.

The rising background of the curves for bias $V \gtrsim 1.2$
corresponds to the feature associated with the main resonance in the
conductance (see the corresponding main central peak in the
conductance curves in Figure \ref{fig:IV_and_IVLandauer_offres}).  
The IETS curve rises at lower bias for the exact calculation compared to
the calculations for the non-interacting system, simply because the
exact calculations include a full renormalization of the electronic
level $\varepsilon_0$. Such a renormalized level is then shifted 
towards lower energy by the full dynamical polaron shift \cite{Dash:2010-I,Ness:2006}.

It is interesting to note that the first-order perturbation correction
to the Landauer current, given by the second term in
Eq.~(\ref{eq:I_Landauer+corrections_withfNEint}) when evaluated from
$G_0$ only, provides not only a qualitatively good feature in the IETS
signal at $V=\omega_0$ but also a partial renormalization of the
electron resonance. That is, the background of the curve rises faster
than for the non-interacting case, and this corresponds to a shift of
the electron resonance towards lower energy by a partial polaron
shift.

Calculations performed non self-consistently provide a partial
renormalization of the electron resonance, however this is closer to
the exact result than that obtained from perturbation theory.
Additionally, the corresponding IETS signal also shows only a feature
at $V=\omega_0$ as for perturbation theory, but its general aspect is
again closer to the exact result.

Another important result of our calculations is that the IETS signal
calculated from only the Landauer-like term in the current $I[G_{BA}]$
and $I[G_{SCBA}]$ does not contain any features at $V=\omega_0$ or
$V=2 \omega_0$, as shown in Figure
\ref{fig:IETS_I_and_ILandauer_offres}.  Although the IETS has the
correct rising background and shows the correct corresponding
renormalization of the electronic level, it fails to correctly
reproduce the features associated with inelastic processes.  
Hence in the off-resonant transport regime the renormalization 
(self-consistent or not) of the \GFs $G^{r,a}$, from which the Landauer-like
current is obtained, does not contain the appropriate physical
information to correctly describe the corresponding IETS signal.
However, as we have seen above, it is good enough to describe the
overall behavior of the conductance at low biases.

Now we turn to the resonant transport regime, and check if the trends 
obtained for the off-resonant regime hold here as well.  
It should be noted that in the
following calculations, we have considered the resonant transport
regime in the case of strong coupling of the central region to the
leads, $t_{0L,R}\sim\beta_{L,R}$.  We are then dealing with an almost
homogeneous one-dimensional system with metallic like behavior at
equilibrium, in which the propagating electrons are coupled locally to
a single localized vibration mode.  We have chosen this somewhat
peculiar regime so that the features in the IETS signal associated
with the inelastic processes are not `distorted' by the features
associated with resonant-like transport. In other words, the
background of the IETS signal around the excitation energies is fairly
flat.

The corresponding IETS curves are shown in Figure
\ref{fig:IETS_I_and_ILandauer_res} for the different kinds of
approximation used to calculate the current. As expected, the IETS
signal calculated for the non-interacting system does not show any
feature at $V=\omega_0$, while the IETS signal obtained from fully
self-consistent calculations show a dip at $V=\omega_0$. Such a
negative contribution to the baseline is to be expected in the case of
good conductors \cite{Galperin:2004b,Frederiksen:2004b} for which
electron-phonon coupling is associated with electron backscattering.
The results obtained from non-self-consistent renormalization are very
similar to the exact results.  Interestingly, the result obtained from
perturbation theory gives a feature in the IETS signal at the right
bias, but however with the wrong sign.

It appears that in all the cases we have studied, first order
perturbation theory always gives a positive contribution (i.e. a peak)
to the IETS which is generally incorrect. Indeed it has been shown that
the inelastic features of the IETS can be both peaks or dips
\cite{Galperin:2004b,Frederiksen:2004b, Paulsson:2005,Egger:2008}
depending on the nature of the conductor and essentially on all the 
parameters describing the system \cite{Egger:2008}.

The most interesting result shown in Figure
\ref{fig:IETS_I_and_ILandauer_res} is that the IETS signal obtained
from only the Landauer-like term in the current $I[G_{BA}]$ and
$I[G_{SCBA}]$ shows the appropriate feature at at $V=\omega_0$. Hence
for the resonant transport regime with strong coupling to the leads,
the renormalization of the \GFs from which the Landauer-like current
is derived is good enough to describe the IETS signature of the
inelastic process at the lowest excitation energy.

\begin{figure}
\begin{center}
\includegraphics[width=\columnwidth]{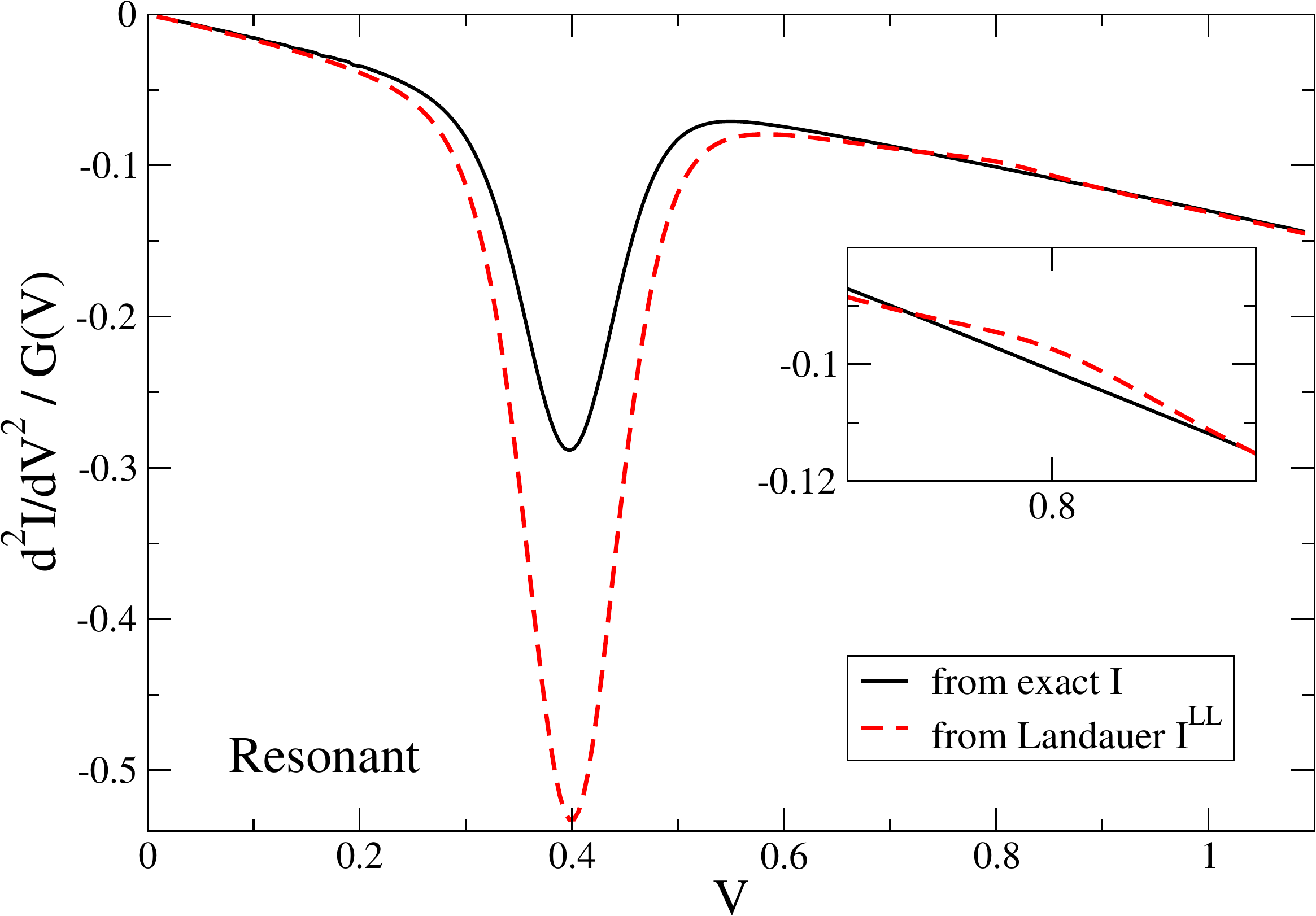}
\end{center}
\caption{IETS signal $d^2I/dV^2$, normalized by $G(V)$, obtained from
  the exact current expression $I(V)$ and from the corresponding
  Landauer-like $I^{\rm LL}(V)$ term.  The \GF calculations are
  performed self-consistently for the resonant case and stronger
  coupling ($\gamma_0/\omega_0=0.875$) than in Figure
  \ref{fig:IETS_I_and_ILandauer_res}.
  Unexpectedly, even for such a strong coupling, there is no
  feature in the IETS signal at $V=2\omega_0$ obtained from the exact
  current. The
  Landauer-like current provides a small feature at $V=2\omega_0$
  with however a spurious positive contribution to the baseline.
  The other parameters are
    $\varepsilon_0=0, \omega_0=0.4, \gamma_0=0.35, t_{0L,R}=1.40,
    T_{L,R}=0.015,
    \eta=0.017, \eta_V=1$.
}
\label{fig:IETS_I_and_ILandauer_res_stronger_evib}
\end{figure}

Finally, we want to comment on the features in the IETS observed at
$V=2\omega_0$, which should correspond to the excitation of two vibron
modes by the injected non-equilibrium charge carriers.  Such a feature
is observed in the rising background of the IETS in the strong
tunneling regime (off-resonant transport regime)
(Fig.~\ref{fig:IETS_I_and_ILandauer_offres}); however this feature is
clearly absent in the resonant transport case.  Even in the case of
strong electron-vibron coupling, shown in Figure
\ref{fig:IETS_I_and_ILandauer_res_stronger_evib}, there is basically
no feature at $V=2\omega_0$ in the IETS signal derived from the exact
expression of the current, although there is a small feature in the
IETS signal derived from the Landauer-like current expression.  The
absence of such a feature at $V=2\omega_0$ for resonant transport is
in agreement with previous studies
(Refs.~[\onlinecite{Frederiksen:2004b,Frederiksen:2007}]). However
at the moment there is no satisfactory physical explanation for the absence 
of such a feature at $V=2\omega_0$.  
We postulate that such an absence may be due to the
partial resummation of the electron-phonon diagrams in the many-body
Green's functions. As we have already shown in Ref.~[\onlinecite{Dash:2010-I}],
higher-order diagrams for the interaction (beyond Hartree-Fock / Born
approximation) play a very important role in correctly describing the
properties of the electron-phonon coupled system, even at intermediate
electron-phonon couplings.

\section{Conclusions}
\label{sec:Conclusions}

In this paper, we have re-addressed the problem of the breakdown
of Landauer-like approaches for electronic transport in the presence of
many-body interactions.
Starting from the original work of Meir and Wingreen \cite{MeirWingreen:1992},
we have once more expressed the current as the sum of a Landauer-like expression 
($I^{\rm LL}$) involving the concept of single-particle transmission probabilities plus a 
non-Landauer-like term arising from the non-equilibrium many-body effects.
We have further developed our theoretical framework to show that the interaction in 
the central scattering region renormalizes not only the non-equilibrium \GFs but also 
the coupling at the contacts between the central region and the leads. 
We have hence obtained a new form of generalized Landauer-like formula for the 
current,  Eqs.(\ref{eq:I_Landauerlike_with_contactrenorm}-\ref{eq:Lambda}), in a 
similar way to
Refs.~[\onlinecite{Sergueev:2002,Zhang:2002,Ferretti:2005a,Ferretti:2005b}].
However our result for the dynamical functional that renormalizes the coupling
at the contacts is more general than the Ng ansatz \cite{Ng:1996} 
used in Refs.~[\onlinecite{Sergueev:2002,Zhang:2002,Ferretti:2005a,Ferretti:2005b}].
Moreover our result does not impose any constraints on the statistical properties of the 
non-equilibrium interacting central region.

We have then applied our theoretical framework to a model system of electron-vibron 
interacting nanojunction. We have analyzed in detail the domain of validity of 
Landauer-like approaches, i.e. without renormalization of the contacts, to describe the conductance
and the inelastic electron tunneling spectroscopy (IETS) of such a non-equilibrium
many-body interacting system.

Our results confirm that generally Landauer-like approaches are not adequate to describe 
the transport properties of such interacting systems for the whole range of applied biases
(linear to highly non-linear regime) and for all the transport regimes (good, metallic-like 
to mediocre, insulator-like conductors). 
In general, the correct transport properties are only obtained from exact non-equilibrium
many-body \GF calculations. 
However, there exist a certain number of conditions in which a Landauer-like approach can 
reproduce fairly well either the conductance or the IETS signal. 
For example, renormalization of the \GFs in a Landauer-like approach is already sufficient 
to account qualitatively for the inelastic features in the IETS signal for the resonant 
transport regime, i.e. $I^{\rm LL}$ gives features at $\omega_0$ in the IETS signal. 
However this is not the case for the off-resonant transport regime,
for which the Landauer-like current $I^{\rm LL}$ fails to reproduce the inelastic features
in the IETS signal.

Finally, we believe that higher order diagrams, as studied in
Ref.~[\onlinecite{Dash:2010-I}], may change the detailed features of
the IETS signal, especially for higher energy excitations.  However
the accuracy to/with which the \GFs are calculated (Born approximation/Hartree-Fock or beyond)
does not alter the main conclusions of our work concerning the applicability
of Landauer-like current formula versus exact derivation of the current.

\begin{acknowledgments}
This work was funded in part by the European Community's Seventh
Framework Programme (FP7/2007-2013) under grant agreement no 211956
(ETSF e-I3 grant).
\end{acknowledgments}

\appendix

\section{Relationships between Green's functions}
\label{App:distribution_fnc}

In order to keep the formalism simple, let us first consider that the
\GFs and the corresponding \SEs are simply complex functions of $\omega$,
i.e. we are dealing with an interacting central region containing only one
site/one electronic level (SSSM model).

When the system is at equilibrium ($f_L=f_R=f^{\rm eq}$),
the lesser (greater) \GF is related to the advanced and retarded \GFs: 
\begin{equation}
\label{eq:Glesser_equi}
G^{<,{\rm eq}}(\omega) = - f^{\rm eq}(\omega)\
\left( G^{r,{\rm eq}}(\omega) - G^{a,{\rm eq}}(\omega) \right),
\end{equation}
and
\begin{equation}
\label{eq:Ggreater_equi}
G^{>,{\rm eq}}(\omega) = - (f^{\rm eq}(\omega)-1)\
\left( G^{r,{\rm eq}}(\omega) - G^{a,{\rm eq}}(\omega) \right). \\
\end{equation}
These relationships are at the center of the fluctuation-dissipation theorem
for equilibrium. They can also be recast as follows,
\begin{equation}
\label{eq:Glessgreat_KMS}
G^{>,{\rm eq}}(\omega)\ =\ - {\rm e}^{(\omega-\mu_0)/kT}\ G^{<,{\rm eq}}(\omega),
\end{equation}
and they then define a relationship between the greater and
lesser Green's functions for statistical averages in the grand canonical ensemble
at finite temperature (the so-called Kubo-Martin-Schwinger boundary conditions 
\cite{Kadanoff:1962,vanLeeuwen:2006}).
Similar relationships also exist for the \SEs $\Sigma^{<,>,r,a}$
(see for example Ref.~[\onlinecite{Jakobs:2010}]).

For non-equilibrium conditions, there is no unique Fermi level at
finite bias (or no unique temperature if $T_L\ne T_R$) in the whole
system, and the relationships given above by Eqs.
(\ref{eq:Glesser_equi}-\ref{eq:Glessgreat_KMS}) no longer hold.  This
is an important feature of the non-equilibrium formalism for which
conventional equilibrium statistics need to be reformulated
\cite{Hershfield:1993}.  However, the self-consistent calculations of
the \GFs and \SEs in the \NE case permit us to define new \NE
distributions.  For example, the non-equilibrium distribution 
$f^{\rm NE}(\omega)$ is defined from the \GFs as follows:
\begin{equation}
\label{eq:fNE}
G^<(\omega) = - f^{\rm NE}(\omega) (G^r(\omega) - G^a(\omega)) .
\end{equation}
This definition is reminiscent of the so-called Kadanoff-Baym ansatz
which has been generalized to the non-equilibrium conditions and to
the time representation of the \GFs 
(see for example Refs. [\onlinecite{Lipavsky:1986,Velicky:2006}]).
Similarly, we also define the non-equilibrium distribution 
$f^{\rm NE}_{\rm int}(\omega)$ from the interaction \SEs as follows:
\begin{equation}
\label{eq:fNEint}
\Sigma^<_{\rm int}(\omega) = - f^{\rm NE}_{\rm int}(\omega) 
( \Sigma^r_{\rm int}(\omega) - \Sigma^a_{\rm int}(\omega) ) .  
\end{equation}

There is no \emph{a priori} reason for these two \NE distribution
functions to be equal to each other at non-equilibrium. 
Both distribution functions contain 
information about both the \NE and the many-body interaction effects 
in the system \cite{Hershfield:1993,NessDashGodby:2010-I}
However, at equilibrium, these distribution functions are as expected
equal to each other and to the conventional Fermi-Dirac equilibrium 
statistics  $f^{\rm NE}=f^{\rm NE}_{\rm int}=f^{\rm eq}$.

As an example, the non-equilibrium distribution function $f^{\rm NE}_0(\omega)$ 
for a non-interacting system \cite{Hershfield:1991}
is given by the weighted averaged of $f_{L,R}$ by the coupling at each
contact $\Gamma_{L,R}$:
\begin{equation}
\label{eq:fNE_0}
f^{\rm NE}_0(\omega)=\frac{ f_L(\omega) \Gamma_L(\omega) + f_R(\omega) \Gamma_R(\omega) } 
{ \Gamma_L(\omega)+\Gamma_R(\omega) } .
\end{equation}
 
The asymptotic values of distribution functions are defined from
behavior of the \GFs and \SEs at large $\omega$, and follow the
conventional statistics: 
\begin{equation}
\begin{split}
& f^{\rm NE}(\omega)=f^{\rm NE}_{\rm int}(\omega)=f^{\rm NE}_0(\omega)
  =f_\alpha(\omega)=f^{\rm eq}(\omega) \\
& \qquad \begin{cases} 
 = 1 , & \omega\rightarrow -\infty \\
 = 0 , & \omega\rightarrow +\infty
\end{cases}
\end{split}
\end{equation}
At equilibrium, one recovers the equilibrium statistics for any distribution function 
$f^{\rm NE}_0=f^{\rm eq}$.

Now, we can generalize our formalism to central regions containing
several electronic states.  The distribution functions then become 
matrices $\mathbf{f}$
\cite{MeirWingreen:1992,Stefanucci:2004a,Stefanucci:2004b} with
elements $f_{nm}$ given by
\begin{equation}
\label{eq:matrix_distribfnc}
X^<_{nm} = - \sum_l f_{nl} ( X^r_{lm} - X^a_{lm}) \ ,
\end{equation}
where $X$ is either a \GF $G$ or a \SE $\Sigma$, and the indices $n,m$ are appropriate
indices to label the electronic states of the central region.

\section{The Ng ansatz}
\label{App:Ng_ansatz}

Here again, we consider in the following mathematical developments
that the \GFs and the \SEs are simply complex functions. Extension to matrices 
is rather straightforward but must be done with care using the notations and definitions 
given in the main text and in Appendix \ref{App:distribution_fnc}.

Using the definition of $f^{\rm NE}_{\rm int}(\omega)$ :
\begin{equation} 
\begin{split}
\Sigma^<_{\rm int}(\omega) & = - f^{\rm NE}_{\rm int}(\omega)\ ( \Sigma^r_{\rm int}(\omega) - \Sigma^a_{\rm int}(\omega)) \\
& = - f^{\rm NE}_{\rm int}(\omega)\ ( \Sigma^>_{\rm int}(\omega) - \Sigma^<_{\rm int}(\omega)),
\end{split}
\end{equation}
and the fact that 
$f_\alpha(\omega)=-{\rm i}\Sigma^<_L(\omega)/\Gamma_L(\omega)$,
we find after more formal manipulations that the renormalization functional
$\Lambda(\omega)$ (Eq.~(\ref{eq:Lambda})) can be re-expressed as:
\begin{equation}
\label{eq:newLambda_1}
\Lambda(\omega)=1+{\rm i}\frac{\Sigma^<_L(\omega) \Sigma^>_{\rm int}(\omega) - \Sigma^>_L(\omega) \Sigma^<_{\rm int}(\omega)}
{\Sigma^<_L(\omega) \Gamma_R(\omega) - \Sigma^<_R(\omega) \Gamma_L(\omega)} \ . 
\end{equation}
After noticing that 
$\Sigma^<_L \Gamma_R - \Sigma^<_R \Gamma_L={\rm i}\Sigma^<_L \Sigma^>_{L+R} - {\rm i} \Sigma^>_L \Sigma^<_{L+R}$, 
we can finally obtain a compact form for $\Lambda(\omega)$:
\begin{equation}
\label{eq:newLambda_2}
\Lambda(\omega)=\frac{\Sigma^<_L(\omega) \Sigma^>(\omega) - \Sigma^>_L(\omega) \Sigma^<(\omega)}
{\Sigma^<_L(\omega) \Sigma^>_{L+R}(\omega) - \Sigma^>_L(\omega) \Sigma^<_{L+R}(\omega)} \ ,
\end{equation}
which is another way of expressing the important result of this paper
given in Eq.~(\ref{eq:Lambda}).

Now we are going to relate our principal results to previous studies using 
the Ng ansatz \cite{Ng:1996,Sergueev:2002,Zhang:2002}.
The Ng ansatz, developed to study the Anderson model out of equilibrium \cite{Ng:1996,
Sergueev:2002,Zhang:2002}, is based on using an apparently more convenient way to 
express the full lesser (greater) \SE in terms of the lesser (greater) \SE for the 
non-interacting system:
\begin{equation}
\label{eq:Ng_ansatz}
\Sigma^{<,>}(\omega)= \Sigma^{<,>}_{L+R}(\omega)\ \bar\Lambda(\omega)
\end{equation}
where $\Sigma^x=\Sigma^x_{L+R}+\Sigma^x_{\rm int}$ and $\Sigma^x_{L+R}=\Sigma^x_L+\Sigma^x_R$,
and
$\bar\Lambda$ is a dynamical ``renormalization'' quantity, to be determined from 
the condition $\Sigma^>-\Sigma^<=\Sigma^r-\Sigma^a$.
We will show below that this ansatz actually implies strong constraints on the statistics
on the non-equilibrium interacting systems.

By using the Ng ansatz to express the lesser and greater \SEs
$\Sigma^{<,>}(\omega)$ in Eq.(\ref{eq:newLambda_2}), one can easily
see that our renormalization functional $\Lambda(\omega)$ given by
Eq.(\ref{eq:newLambda_2}) is just equal to the dynamical quantity
$\bar\Lambda(\omega)$ of the Ng ansatz:
$\Lambda(\omega)=\bar\Lambda(\omega)$.  We then recover all the
expressions for the current previously derived in
Refs.~[\onlinecite{Ng:1996, Sergueev:2002, Ferretti:2005a,
  Ferretti:2005b, Zhang:2002}] from our main results Eqs.
(\ref{eq:I_Landauerlike_with_contactrenorm}-\ref{eq:Lambda}).

However, the Ng ansatz presents some intrinsic limitations. To prove
this, it is sufficient to calculate the \NE distribution function
$f^{\rm NE}_{\rm int}$ within the Ng ansatz.  Starting from the
definition of $f^{\rm NE}_{\rm int}(\omega)$, i.e.
\begin{equation}
\label{eq:fNEint_appendix_1}
f^{\rm NE}_{\rm int}(\omega) = - \Sigma^<_{\rm int}(\omega) / ( \Sigma^r_{\rm int}(\omega) - \Sigma^a_{\rm int}(\omega)) \ ,
\end{equation}
it is straightforward to show that $f^{\rm NE}_{\rm int}$ is then given by
\begin{equation}
\label{eq:fNEint_appendix_2}
\begin{split}
 f^{\rm NE}_{\rm int}(\omega) & = 
- \frac{ \Sigma^{<}_{L+R}(\omega) (\bar\Lambda(\omega)-1) }
{ (\Sigma^{>}_{L+R} - \Sigma^{<}_{L+R}) (\bar\Lambda-1) } \\   
& = \frac{ f_L(\omega) \Gamma_L(\omega)+f_R(\omega) \Gamma_R(\omega)}{\Gamma_L(\omega)+\Gamma_R(\omega)}
= f^{\rm NE}_0(\omega)  , 
\end{split}
\end{equation}
the \NE 
distribution function for the non-interacting system! 

Just to confirm the consistency of our derivations, if we use the
above result $f^{\rm NE}_{\rm int}(\omega)=f^{\rm NE}_0(\omega)$ in
the definition of our renormalization functional $\Lambda(\omega)$
given by Eq. (\ref{eq:Lambda}) and the Ng ansatz for
$\Sigma^{<,>}_{\rm int}$, i.e. $\Sigma^{<,>}_{\rm
  int}(\omega)=\Sigma^{<,>}_{L+R}(\omega)\ (\bar\Lambda(\omega)-1)$,
we find again and consistently that
$\Lambda(\omega)=1+(\bar\Lambda(\omega)-1)=\bar\Lambda(\omega)$, as
expected.

However we have found \cite{NessDashGodby:2010-I} that the condition
$f^{\rm NE}_{\rm int}(\omega)=f^{\rm NE}_0(\omega)$ implies
necessarily that $f^{\rm NE}(\omega)=f^{\rm NE}_0(\omega)$.  In other
terms, the Ng ansatz implies that the full \NE distribution $f^{\rm
  NE}$ of the interacting system, as well as $f^{\rm NE}_{\rm int}$,
are equal to the \NE distribution function for the non-interacting
system! This is a condition that is in contradiction with the fact that both
distribution functions should simultaneously include  both the \NE effects
and the many-body interaction effects.  In fact $f^{\rm NE}(\omega)$
is actually a functional of both the \NE distribution function for the
non-interacting $f^{\rm NE}_0(\omega)$ and the many-body interaction
\cite{Hershfield:1993,NessDashGodby:2010-I}.

Hence we conclude that our expression for the renormalization of the
coupling at the contact $\Lambda(\omega)$ given by
Eq.~(\ref{eq:Lambda}) is more general than the definition used in the
Ng ansatz.  The latter is not taking fully into account the
interaction effects at non equilibrium.  It actually corresponds to a
lowest-order expansion of the full \NE distribution in terms of only
the \NE distribution function of the non-interacting system
\cite{NessDashGodby:2010-I}.  Or in other words, the Ng ansatz
considers that the statistics of the interacting central region is
dominated by that of the non-interacting leads at non-equilibrium, and
that the interaction effects in the central region do not affect its
\NE statistics.

Finally we would like to mention that it is however possible to
recover the Ng ansatz from our results in the limit of low-energy
scales \cite{NessDashGodby:2010-I}, i.e.  when
$(\omega-\mu_\alpha)\sim 0$ and then $\exp (\omega-\mu_\alpha)/kT \sim
1$.  This implies that although approximate, the Ng ansatz might be
good enough to describe low-energy excitations like, for example, the
Kondo effect in correlated electron systems, which gives a sharp
feature in the spectral density around the Fermi level at equilibrium,
or split Kondo peaks around the leads' Fermi levels at non
equilibrium \cite{Meir:1993}.  However, such an ansatz will most
probably fail to describe systems for which the interaction
(electron-phonon, electron-plasmon) is restricted on an energy scale
defined by the phonon (plasmon) frequency, which is finite and not
necessarily small \cite{Dash:2010-I,NessDashGodby:2010-I}.

\end{document}